\documentclass[conference]{IEEEtran}
\IEEEoverridecommandlockouts

\usepackage{cite}
\usepackage{amsmath,amssymb,amsfonts,bm}
\usepackage{algorithmic}
\usepackage{graphicx}
\usepackage{textcomp}
\usepackage{xcolor}

\usepackage{multirow}
\usepackage{siunitx}
\usepackage{url}

\def\BibTeX{{\rm B\kern-.05em{\sc i\kern-.025em b}\kern-.08em
    T\kern-.1667em\lower.7ex\hbox{E}\kern-.125emX}}

\def\transpose{\mathsf{T}}
\def\R{\mathbb{R}}

\def\AudioEnc{\mathrm{Enc}}
\def\TagDec{\mathrm{Dec}}
\def\ChainTagDec{\mathrm{ChainDec}}
\def\AffineTransform{\mathrm{Affine}}
\def\GRU{\mathrm{GRU}}
\def\Concat{\mathrm{Concat}}

\def\g{\mathrm{g}}
\def\i{\mathrm{i}}
\def\m{\mathrm{m}}
\def\gim{\g\rightarrow\i\rightarrow\m}
\def\gmi{\g\rightarrow\m\rightarrow\i}
\def\igm{\i\rightarrow\g\rightarrow\m}
\def\img{\i\rightarrow\m\rightarrow\g}
\def\mgi{\m\rightarrow\g\rightarrow\i}
\def\mig{\m\rightarrow\i\rightarrow\g}

\begin{document}

\title{Music Tagging with Classifier Group Chains}

\author{\IEEEauthorblockN{Takuya Hasumi}
\IEEEauthorblockA{\textit{LY Corporation} \\
Tokyo, Japan \\
takuya.hasumi@lycorp.co.jp}
\and
\IEEEauthorblockN{Tatsuya Komatsu}
\IEEEauthorblockA{\textit{LY Corporation} \\
Tokyo, Japan \\
komatsu.tatsuya@lycorp.co.jp}
\and
\IEEEauthorblockN{Yusuke Fujita}
\IEEEauthorblockA{\textit{LY Corporation} \\
Tokyo, Japan \\
yusuke.fujita@lycorp.co.jp}
}

\maketitle

\begin{abstract}
We propose music tagging with classifier chains that model the interplay of music tags.
Most conventional methods estimate multiple tags independently by treating them as multiple independent binary classification problems.
This treatment overlooks the conditional dependencies among music tags, leading to suboptimal tagging performance.
Unlike most music taggers, the proposed method sequentially estimates each tag based on the idea of the classifier chains.
Beyond the naive classifier chains, the proposed method groups the multiple tags by category, such as genre, and performs chains by unit of groups, which we call \textit{classifier group chains}.
Our method allows the modeling of the dependence between tag groups.
We evaluate the effectiveness of the proposed method for music tagging performance through music tagging experiments using the MTG-Jamendo dataset.
Furthermore, we investigate the effective order of chains for music tagging.
\end{abstract}

\begin{IEEEkeywords}
Music tagging, multi-label classification, chain rule, classifier chains
\end{IEEEkeywords}

\section{Introduction}
\label{sec:introduction}
Music tagging~\cite{minz2021music} is the task of estimating the attributes of a given music segment, such as genres, moods, and instruments.
The estimated attributes, called tags, enable efficient searching within large-scale music databases.
The music tracks searched by the tags are then used to build mood-based playlists and genre-specific radio stations.
Accurate music tagging is crucial for music streaming services to improve the user experience.

To improve the performance of the music tagging system, many deep-learning-based methods~\cite{choi2016automatic,lee2017sample,pons2018end,kim2018sample,choi2019zero,won2019toward,won2020evaluation,won2021semi,manco2022learning,mccallum2022supervised,doh2023toward} have been proposed so far.
Recent music tagging studies mainly focus on improving performance using semi-supervised and self-supervised learning.
These studies show the effectiveness of using a large number of music tracks collected from the Web or proprietary databases.
In these studies (e.g., \cite{doh2023toward}), the tagging system is trained by the framework of linear probing.
The encoder of the tagging system is pretrained by the large dataset, and the decoder is then fine-tuned by a simple module such as a stack of affine transformation and sigmoid function.
While these studies enhance the feature extraction ability of the encoder based on the linear probing framework, they overlook the improvement of the decoder.
Specifically, these methods estimate multiple music tags in parallel by treating them as flattened estimation targets.
Since the parallel decoder architecture assumes conditional independence of the output tags, these methods cannot consider the hierarchy and interplay between music tags.

Some prior works~\cite{garcia2021leveraging, zhi2023attention} have proposed using hierarchical dependency in the music instrument recognition task~\cite{krishna2004music} to handle the group structure of the instrument tags.
In these works, they treat the hierarchy of instruments, such as the relationship of Brass (top) and Trombone (bottom).
Using the top coarse and bottom fine labels, they trained instrument recognition models by multi-task learning.
While their works can capture the coarse-to-fine hierarchy of music tags, they do not address the interplay between the labels or instrument groups.

For capturing the interplay between different labels in multi-label classification, classifier-chain approaches~\cite{read2011classifier,cheng2010bayes} have been investigated in related fields such as source separation~\cite{shi2020speaker}, speaker diarization~\cite{fujita2020neural,shi2020sequence}, and sound event detection~\cite{komatsu2021acoustic}, where there is arbitrariness in the order of handling multi-audio events.
These methods overcome the limitation of assuming that sound events are mutually independent by considering their dependencies, which results in higher performance.
However, these classifier chains have been utilized with several label classes --- at most ten in \cite{komatsu2021acoustic}.
Furthermore, these studies did not consider any group structure among the labels.
Their approaches are not directly applicable to music tagging, which sometimes involves hundreds of label classes organized in a hierarchical structure.

In this paper, we propose music tagging with classifier chains~\cite{read2011classifier} by applying the chain rule to the music tagging problem.
To consider the hierarchical structure of the music tags, we propose a category-based group chain that determines the order of the tag group estimation, called \textit{classifier group chains}.
The proposed music tagger sequentially estimates the music tags in one category group, e.g., genres, conditioned on the previous estimations of different category groups, such as instruments and moods.
As a result, the proposed method can capture the relationship between different musical attributes.
We demonstrate the effectiveness of our proposed method through music tagging experiments using MTG-Jamendo~\cite{bogdanov2019mtg} dataset, which indicates the necessity to consider the dependence between different aspects of the music piece in the music tagging task.

\section{Conventional music tagging}
\label{sec:conventional-music-tagging}
\begin{figure}[tb]
    \centering
    \includegraphics[width=0.85\linewidth]{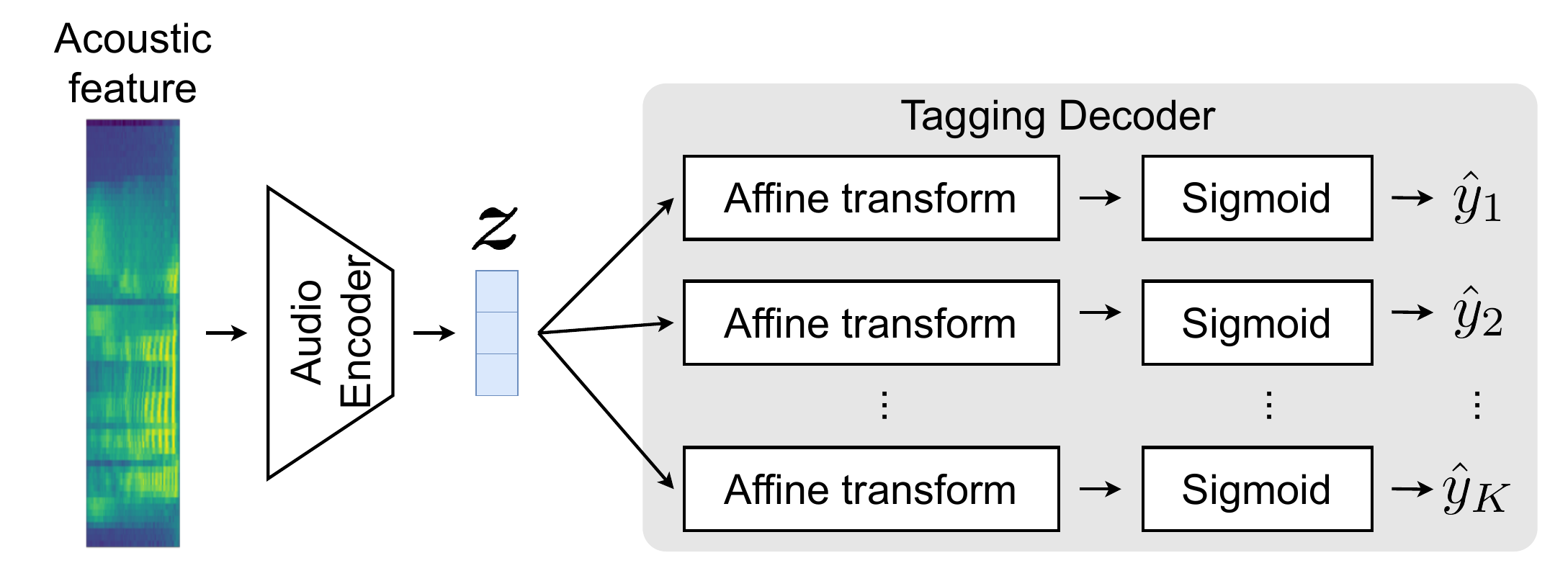}
    \caption{Overview of conventional music tagging system with $K$ music tags. $K$ sub-decoders (affine transformation + sigmoid function) estimate each binary flag of the music tag independently.}
    \label{fig:conventional-music-tagging/overview}
\end{figure}

Music tagging is treated as a multi-label classification problem~\cite{minz2021music} with $K$ tags.
Fig.~\ref{fig:conventional-music-tagging/overview} shows the overview of the general conventional method for music tagging with $K$ tags.
In the conventional methods, an audio encoder $\AudioEnc$ transforms temporal acoustic features $\bm{a}$ (e.g., Mel-spectrogram) into temporally aggregated latent feature vector $\bm{z}\in\R^{D}$, where $D$ denotes the number of dimensions.
\begin{align}
    \bm{z} = \AudioEnc(\bm{a}).
\end{align}
By feeding $\bm{z}$ to a tagging decoder $\TagDec$, it outputs the $K$ probabilities denoted as $\hat{\bm{y}}=(\hat{y}_{1},\ldots,\hat{y}_{K})^{\transpose}\in[0,1]^{K}$.
\begin{align}
    \hat{\bm{y}} = \TagDec(\bm{z}).
\end{align}
The conventional $\TagDec$ can be described as $K$ sub-decoders, each of which independently estimates one of $K$ tags.
\begin{align}
    \hat{y}_{k} = \TagDec^{(k)}(\bm{z}),
\end{align}
where $k$ denotes the index of $k$th music tag.
Typically, $\TagDec^{(k)}$ is implemented as follows:
\begin{align}
    \TagDec^{(k)}(\bm{z})
    &= \sigma(\AffineTransform(\bm{z})),
\end{align}
where $\AffineTransform(\cdot)$ and $\sigma(\cdot)$ denote the affine transformation and sigmoid function, respectively.
In this modeling, we assume the following probabilistic model during inference:
\begin{align}
    p(\hat{\bm{y}}|\bm{z})
    &\approx \prod_{k}p(\hat{y}_{k}|\bm{z}).
    \label{eq:conventional-music-tagging/probabilistic-chains}
\end{align}
Here, the approximation in Eq.~\eqref{eq:conventional-music-tagging/probabilistic-chains} implies the assumption that the music tags are conditionally independent by the latent feature $\bm{z}$.

\section{Music tagging with classifier group chains}
\label{sec:music-tagging-with-classifier-chains}
\subsection{Motivation}
\label{sec:music-tagging-with-classifier-chains/motivation}
Although the music tags contain some structure, such as tag categories, conventional music tagging methods treat all tags equally and independently.
This treatment does not capture the group structure of the music tags.
Considering such a structure of multiple tags is intuitively understandable.
For example, by observing the music instrument tags ``guitar'', ``electronic piano'', and ``drums'', we can empirically estimate that music can be labeled as ``pop'' and cannot be labeled as ``classical'' for music genre tags.
Therefore, conditional inference based on the structure of music tags may enhance the performance of the music tagging thanks to its context-aware predictions.

\subsection{Formulation}
\label{sec:music-tagging-with-classifier-chains/formulation}
\begin{figure}
    \centering
    \includegraphics[width=0.98\linewidth]{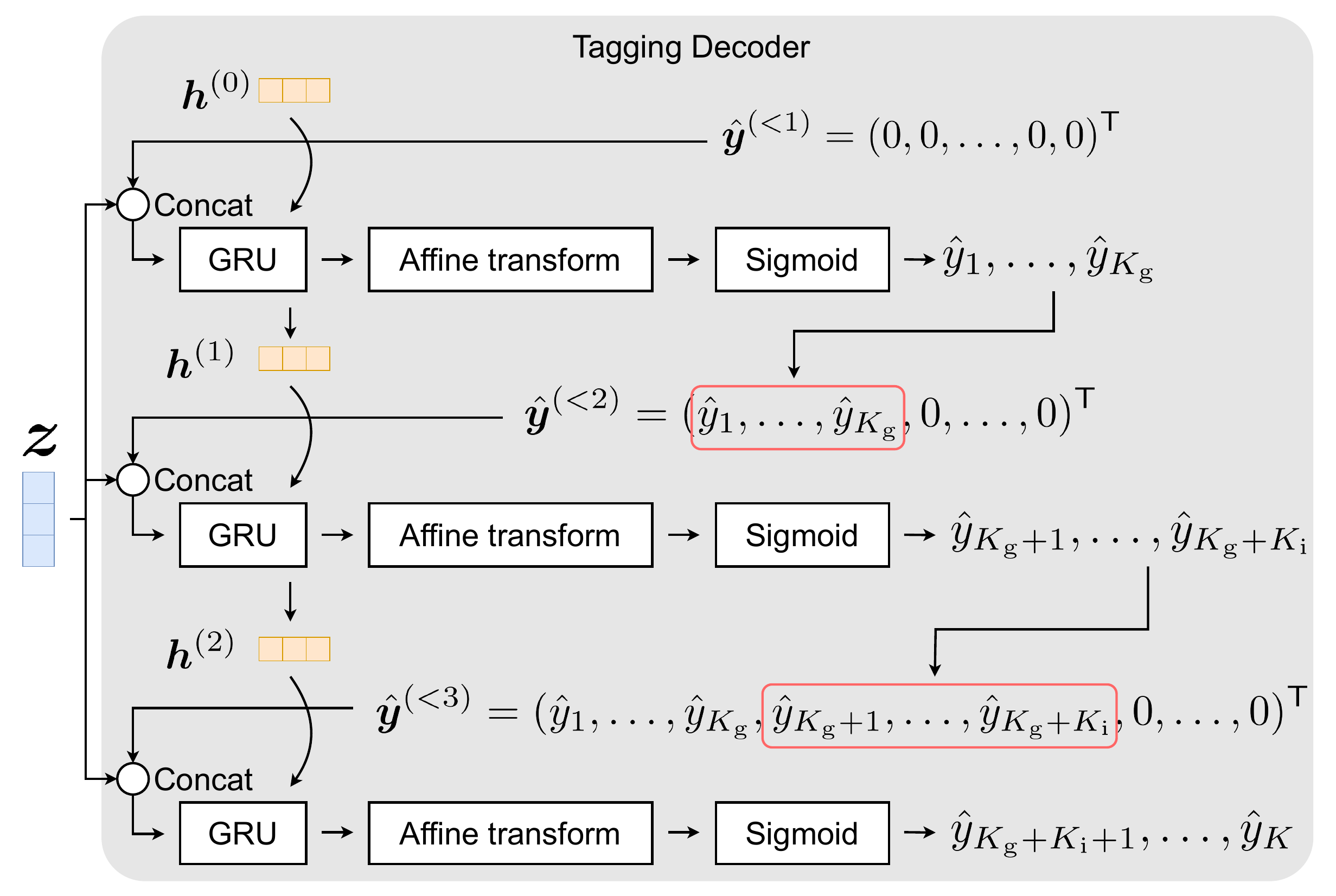}
    \caption{Overview of the proposed music tagging with classifier group chains ``genre'' $\rightarrow$ ``instrument'' $\rightarrow$ ``mood/theme''. Only the decoder part is shown, and the encoder part is identical to the one in Fig.~\ref{fig:conventional-music-tagging/overview}. Unlike the conventional method, $\gamma$th music tag group (category) is sequentially estimated using the previous estimation results.}
    \label{fig:music-tagging-with-classifier-chains/group-chains/overview-of-group-chained-decoder}
\end{figure}
To capture the dependencies of the group structure in the music tags, which are not considered in the conventional tagging decoders, we propose a model that handles the interplay of music tags inspired by the classifier chains~\cite{read2011classifier}.

In the proposed music tagging with classifier group chains, we split tags into multiple groups based on tag category.
Specifically, we split them into three categories, each reflecting different aspects of the music piece: genre, instrument, and mood/theme.
Here, we consider the chain of ``genre'' $\rightarrow$ ``instrument'' $\rightarrow$ ``mood/theme'', and set $\g=1$, $\i=2$, $\m=3$ for simplicity.
Furthermore, we assume the first $K_{\g}$ tags are associated with genre, the second $K_{\i}$ tags are associated with instruments, and the third $K_{\m}$ tags are associated with mood or theme satisfying $K = K_{\g} + K_{\i} + K_{\m}$.
Using these groups, we split $\hat{\bm{y}}$ into $\hat{\bm{y}}^{(\g)}\in [0, 1]^{K_{\g}}$, $\hat{\bm{y}}^{(\i)}\in [0, 1]^{K_{\i}}$, and $\hat{\bm{y}}^{(\m)}\in [0, 1]^{K_{\m}}$.
We decompose $p(\hat{\bm{y}}|\bm{z})$ as follows:
\begin{align}
    p(\hat{\bm{y}}|\bm{z})
    &= p(\hat{\bm{y}}^{(\m)}|\hat{\bm{y}}^{(\i)},\hat{\bm{y}}^{(\g)},\bm{z})p(\hat{\bm{y}}^{(\i)}|\hat{\bm{y}}^{(\g)},\bm{z})p(\hat{\bm{y}}^{(\g)}|\bm{z}).
    \label{eq:music-tagging-with-classifier-chains/group-chains}
\end{align}
To model Eq.~\eqref{eq:music-tagging-with-classifier-chains/group-chains}, we sequentially estimate tag groups one by one.
In the proposed method, the latent feature vector $\bm{z}$ is fed to the tagging decoder, which recurrently estimates each tag group based on prior tag group estimation.
\begin{align}
    \hat{\bm{y}}^{(\gamma)}, \bm{h}^{(\gamma)} = \ChainTagDec^{(\gamma)}(\bm{z},\hat{\bm{y}}^{(<\gamma)}, \bm{h}^{(\gamma-1)}).
\end{align}
$\hat{\bm{y}}^{(<\gamma)}\in [0, 1]^{K}$ denotes the vector, where $1$st through $K_{1} + \ldots + K_{\gamma-1}$ elements are filled with the previous estimations at ${\gamma}$th chain and the remaining entries are filled with $0$.
By collecting $\hat{\bm{y}}^{(\gamma)}$, we obtain the final estimation of all tags.

Unlike the tagging decoder used in conventional methods, the one used in the proposed method iteratively takes latent feature vector $\bm{z}$, previously estimated tag groups $\hat{\bm{y}}^{(<\gamma)}$, and previous hidden state $\bm{h}^{(\gamma-1)}$.
To manage the state in iterative inference, we employ a gated recurrent unit (GRU)~\cite{chung2014empirical}, one of the recurrent neural networks (RNNs), as a backbone of the tagging decoder.
In this architecture, the tagging decoder for the $\gamma$th tag group is written as:
\begin{align}
    \ChainTagDec^{(\gamma)}(\bm{z},\hat{\bm{y}}^{(<\gamma)}, \bm{h}^{(\gamma-1)})
    &= (\sigma(\AffineTransform^{(\gamma)}(\bm{\eta})),~\bm{h}^{(\gamma)}), \\
    \bm{\eta}, \bm{h}^{(\gamma)}
    &= \GRU(\bm{\zeta}, \bm{h}^{(\gamma-1)}), \\
    \bm{\zeta}
    &= \Concat(\bm{z},\hat{\bm{y}}^{(<\gamma)}),
\end{align}
where $\bm{h}^{(\gamma)}$ corresponds to a hidden cell of a GRU and $\Concat(\cdot)$ represents the concatenation to yield a supervector of $\bm{z}$ and $\hat{\bm{y}}^{(<\gamma)}$.

\subsection{Relation to classifier chains}
\label{sec:music-tagging-with-classifier-chains/relation-to-classifier-chains}
The concept of classifier group chains is inspired by the idea of the classifier chains~\cite{read2011classifier}.
In the naive classifier chains, $p(\hat{\bm{y}}|\bm{z})$ is modeled as
\begin{align}
    p(\hat{\bm{y}}|\bm{z})
    = \prod_{k}p(\hat{y}_{k}|\hat{y}_{1},\ldots,\hat{y}_{k-1},\bm{z}).
    \label{eq:music-tagging-with-classifier-chains/relation-to-classifier-chains/chains}
\end{align}
However, when we apply the naive classifier chains to the music tagging problem, multiple options exist for the order of tags.
The popular music tagging dataset MTG-Jamendo~\cite{bogdanov2019mtg} contains at least 50 tags.
Even in this most minor case, $50! \approx 3\times 10^{64}$ patterns can be executable, which makes finding the best order of chains unrealistic.

By grouping the multiple tags by tag category as described in Sec.~\ref{sec:music-tagging-with-classifier-chains/formulation}, we can significantly reduce the number of possible patterns to search for.
Furthermore, this grouping allows the classifier chains to capture the relationships within and between different categories of tags.
By structuring the tags in this manner, we aim to enhance the ability of the model to make more contextually relevant predictions.

\section{Experiments}
\label{sec:experiments}
\begin{table}[tb]
    \caption{Member samples of each category in MTG-Jamendo dataset. We show only the five tags and remove the prefix text defined in the original dataset due to space limitations.}
    \begin{center}
    \begin{tabular}{c|ccc}
    \hline
        & Genre & Instrument & Mood/Theme  \\ \hline \hline
        & alternative & acousticguitar & emotional \\
        & ambient & bass & energetic \\
        Examples & atmospheric & computer & film \\
        & chillout & drummachine & happy \\
        & classical & drums & relaxing \\ \hline
        Number of tags & \multirow{2}{*}{$31$~/~$87$} & \multirow{2}{*}{$14$~/~$40$} & \multirow{2}{*}{$5$~/~$56$} \\
        \texttt{top50}~/~\texttt{all} &  &  &  \\\hline
    \end{tabular}
    \label{tab:experiments/dataset/category-members}
    \end{center}
\end{table}
\begin{table*}[tb]
    \caption{Comparison of music tagging performance using \texttt{top50} subset in MTG-Jamendo dataset. $\g$, $\i$, and $\m$ are shorts of genre, instrument, mood/theme, respectively.}
    \label{tab:experiments/results/mtg-top50}
    \begin{center}
    \begin{tabular}{c|c|cccccc|cc}
        \hline
        \multirow{2}{*}{Tagging decoder} & \multirow{2}{*}{Order of chain} & \multicolumn{2}{c}{Genre} & \multicolumn{2}{c}{Instrument} & \multicolumn{2}{c}{Mood/Theme} & \multicolumn{2}{|c}{All} \\
         & & ROC-AUC & PR-AUC & ROC-AUC & PR-AUC & ROC-AUC & PR-AUC & ROC-AUC & PR-AUC \\ \hline
        \multicolumn{1}{l|}{\textit{Baselines}} &  &  & & & & & & & \\
        Affine & --- & $85.7$ & $32.8$ & $76.0$ & $19.3$ & $73.6$ & $10.8$ & $81.8$ & $27.0$ \\
        GRU & --- & $85.9$ & $33.8$ & $\bm{76.6}$ & $\bm{19.9}$ & $74.2$ & $11.5$ & $82.1$ & $\bm{27.6}$ \\
        \multicolumn{1}{l|}{\textit{Proposed}} &  &  & & & & & & & \\
        GRU & $\gim$ & $86.0$ & $33.5$ & $76.5$ & $\bm{19.9}$ & $74.0$ & $\bm{11.7}$ & $82.1$ & $27.5$ \\
        GRU & $\gmi$ & $\bm{86.1}$ & $\bm{33.9}$ & $76.0$ & $19.6$ & $73.3$ & $11.2$ & $82.0$ & $\bm{27.6}$ \\
        GRU & $\igm$ & $86.0$ & $33.6$ & $76.5$ & $\bm{19.9}$ & $\bm{74.6}$ & $11.5$ & $\bm{82.2}$ & $\bm{27.6}$ \\
        GRU & $\img$ & $85.9$ & $33.4$ & $76.5$ & $\bm{19.9}$ & $74.4$ & $11.6$ & $82.1$ & $27.5$ \\
        GRU & $\mgi$ & $85.9$ & $33.5$ & $76.2$ & $19.6$ & $74.1$ & $11.4$ & $82.0$ & $27.4$ \\
        GRU & $\mig$ & $86.0$ & $33.7$ & $\bm{76.6}$ & $19.8$ & $73.6$ & $10.8$ & $\bm{82.2}$ & $27.5$ \\ \hline
    \end{tabular}
    \end{center}
\end{table*}
\begin{table*}[tb]
    \caption{Comparison of music tagging performance using \texttt{all} subset in MTG-Jamendo dataset.}
    \label{tab:experiments/results/mtg-all}
    \begin{center}
    \begin{tabular}{c|c|cccccc|cc}
        \hline
        \multirow{2}{*}{Tagging decoder} & \multirow{2}{*}{Order of chain} & \multicolumn{2}{c}{Genre} & \multicolumn{2}{c}{Instrument} & \multicolumn{2}{c}{Mood/Theme} & \multicolumn{2}{|c}{All} \\
         & & ROC-AUC & PR-AUC & ROC-AUC & PR-AUC & ROC-AUC & PR-AUC & ROC-AUC & PR-AUC \\ \hline
         \multicolumn{1}{l|}{\textit{Baselines}} &  &  & & & & & & & \\
        Affine & --- & $85.3$ & $16.5$ & $\bm{76.0}$ & $\bm{10.4}$ & $75.2$ & $6.6$ & $80.2$ & $12.1$ \\
        GRU & --- & $85.3$ & $16.1$ & $74.0$ & $10.0$ & $76.0$ & $6.5$ & $79.9$ & $11.8$ \\
        \multicolumn{1}{l|}{\textit{Proposed}} &  &  & & & & & & & \\
        GRU & $\gim$ & $85.9$ & $\bm{17.0}$ & $74.3$ & $10.1$ & $76.2$ & $6.8$ & $80.4$ & $\bm{12.4}$ \\
        GRU & $\gmi$ & $85.8$ & $16.6$ & $73.8$ & $10.0$ & $75.4$ & $6.7$ & $80.0$ & $12.1$ \\
        GRU & $\igm$ & $85.6$ & $16.3$ & $75.1$ & $10.2$ & $76.1$ & $6.5$ & $80.4$ & $12.0$ \\
        GRU & $\img$ & $\bm{86.0}$ & $16.6$ & $75.3$ & $10.3$ & $77.0$ & $7.0$ & $\bm{80.9}$ & $12.3$  \\
        GRU & $\mgi$ & $85.7$ & $16.5$ & $74.8$ & $10.0$ & $77.1$ & $7.2$ & $80.7$ & $12.2$ \\
        GRU & $\mig$ & $85.9$ & $16.5$ & $75.0$ & $10.1$ & $\bm{77.5}$ & $\bm{7.3}$ & $\bm{80.9}$ & $12.3$ \\ \hline
    \end{tabular}
    \end{center}
\end{table*}
\subsection{Datasets}
\label{sec:experiments/datasets}
To evaluate the effectiveness of the proposed method, we performed music tagging experiments using the MTG-Jamendo dataset~\cite{bogdanov2019mtg}.
The MTG-Jamendo dataset includes about 200 tags in total, and each tag can be officially categorized as ``genre'', ``instrument'', or ``mood/theme''.
We used this official categorization for classifier group chains.
To include all categories in the training dataset, we chose two variants of the MTG-Jamendo dataset.
One is a \texttt{top50}, which includes the frequently annotated 50 tags and excludes others.
The other is \texttt{all}, including 183 tags.
We chose the \texttt{split-0} for each set, which is generally used to compare the music taggers.
Table \ref{tab:experiments/dataset/category-members} shows the examples and number of samples in each category.

We downsampled all the music tracks into $16\si{\kilo\Hz}$.
Each clip is transformed into $128$-dimensional Mel-spectrogram by $25\si{\milli\second}$ window and $10\si{\milli\second}$ hop length, resulting in $128\times 1000$ acoustic features $\bm{a}$.
During training, we randomly selected the $10$-second audio clip and applied SpecAugment~\cite{park2019specaugment} to the acoustic features following~\cite{koutini2022efficient}.
During the validation and evaluation, eight equally spaced $10$-second clips were selected, and the final estimation was aggregated by averaging the outputs of each clip.

\subsection{Model and training details}
We compared the music tagging models by changing the tagging decoders.
As an audio encoder for all tagging models, we used the pre-trained patchout fast spectrogram
Transformer (PaSST)~\cite{koutini2022efficient}.
The architecture of PaSST is based on the vision Transformer (ViT)~\cite{dosovitskiy2020image,touvron2021training}, which enables efficient feature extraction and faster training by disentangled positional encodings and patchout.
We downloaded the publicly available model \texttt{passt\_s\_swa\_p16\_128\_ap476}\footnote{\url{https://github.com/kkoutini/PaSST}} pretrained by the AudioSet~\cite{gemmeke2017audio}.

As a tagging decoder of the simple baseline, we used the cascade of affine transformation and sigmoid function as described in Sec.~\ref{sec:conventional-music-tagging}.
In addition, following \cite{won2021semi}, we prepend layer normalization~\cite{ba2016layer} before affine transformation.
In the proposed method, we used GRUs~\cite{chung2014empirical} to estimate tags iteratively as described in Sec.~\ref{sec:music-tagging-with-classifier-chains/formulation}.
The number of GRU layers is $1$, and the number of hidden sizes is $128$.
The initial hidden state is set to a trainable parameter.
To investigate the advantage of the GRU, we also compared the GRU without classifier chain as a tagging decoder.
We treat the GRU-based tagger without classifier group chains as one of the baseline methods.

For both baseline and proposed methods, we froze the audio encoder and only fine-tuned the tagging decoder following the conventional framework of music tagging.
We trained the models with baseline and the proposed decoders with a batch size of $128$ for $100$ epochs.
For optimizing the parameters, we used Adam~\cite{kingma2014adam} with a learning rate of $0.0001$.
For evaluation, we choose the best model based on the loss for the validation dataset.

\subsection{Results}
\label{sec:experiments/results}
We evaluated the baseline and proposed methods by computing the area
under receiver operating characteristic curve (AUC-ROC) and the area under precision-recall curve (PR-ROC) to each category and the entire tag.
Table~\ref{tab:experiments/results/mtg-top50} and \ref{tab:experiments/results/mtg-all} show the comparison of the baseline and proposed methods' music tagging performance for \texttt{top50} and \texttt{all}, respectively.

From the results, while using the GRU improves the overall scores in \texttt{top50}, it does not improve the ones in \texttt{all}.
However, the proposed music tagging with classifier chains improved the overall ROC- and PR-AUC scores in most orders of chains.
Based on these observations, the GRU architecture only contributes to limited performance improvement for the proposed methods.
Furthermore, for any tag group, the proposed methods show higher scores in the best settings except for the scores of ``instrument'' in \texttt{all}.
These improvements show that chaining the tags or tag groups enhances the capture of the conditional dependence between different tag categories.

\subsection{Effect of group chain order}
Table~\ref{tab:experiments/results/mtg-top50} and \ref{tab:experiments/results/mtg-all} also show the importance of the order of chains.
Regarding the tagging performance for each category, the case where the target category is first estimated shows relatively higher scores for ``genre'' in \texttt{top50}.
At first glance, it seems that the classifier group chains have no advantage since the first category shows the highest score.
However, during the training of the tagging decoder with classifier group chains, the estimation of the first category is passed to the next sub-decoder without detaching the gradient.
Therefore, compared to the later categories in chains, the estimation of the first categories can be optimized using both the first and subsequent estimations, which increases its performance over the other categories.
However, this interpretation is limited rather than the results of ``genre'' in \texttt{top50} and ``mood/theme'' in \texttt{all}.
For example, the scores of ``instrument'' in \texttt{top50} show relatively higher scores when ``instrument'' is estimated to be first or second.
In the case of ``mood/theme'' $\rightarrow$ ``instrument'' $\rightarrow$ ``genre'' in \texttt{top50}, even though ``mood/theme'' is first estimated, the scores for the category are relatively lower.
This can be due to the dominance of genre estimation in \texttt{top50}.
In this task, $62$\% tags are categorized as ``genre'' as shown in Table.~\ref{tab:experiments/dataset/category-members}\footnote{$31/(31 + 14 + 5) \approx 0.62$}.
It is also possible to prioritize the optimization of dominant category ``genre'' at the expense of the performance of less frequent categories to minimize the loss during the training.
By increasing the ratio of the target category (e.g., ``mood/theme'' in \texttt{all}),  the score for the category becomes higher than estimated later.

\section{Conclusion}
\label{sec:conclusion}
This paper proposed the music tagger with classifier group chains, a novel approach to more accurately estimating multiple music tags.
Unlike conventional music taggers, our method sequentially estimates music tags per category based on chain rules classifiers, which enables the modeling of the conditional dependence between tags.
Experimental results using the MTG-Jamendo dataset showed that our method outperforms conventional methods, which treat each tag independently.
Furthermore, we experimentally showed that the order of chains affects tagging performance in each category.
For future work, we plan to compare and integrate the music captioning approaches~\cite{cai2020music,doh2023lp}, which implicitly capture the dependency among tags using the language model.

\bibliographystyle{IEEEbib}

\end{document}